\documentclass[aps,twocolumn,showpacs,showkeys,preprintnumbers,amsmath,amssymb,superscriptaddress,nofootinbib]{revtex4-1}
\usepackage{graphicx}
\usepackage{color}
\usepackage{hyperref}
\def\ps@pprintTitle{%
\let\@oddhead\@empty
\let\@evenhead\@empty
\def\@oddfoot{}%
\let\@evenfoot\@oddfoot}
\def\GeV2 {$\mathrm{GeV} ^2$}

\def\mpm {$\pm$}
\def\dsigel0 {$d\sigma_{el} /dt \vert _{t=0}$}

\begin{document}

\title{Results on Total and Elastic Cross Sections in Proton--Proton Collisions at $\sqrt{s} = 200$~GeV}

%
%
\affiliation{Abilene Christian University, Abilene, Texas   79699}
\affiliation{AGH University of Science and Technology, FPACS, Cracow 30-059, Poland}
\affiliation{Alikhanov Institute for Theoretical and Experimental Physics NRC "Kurchatov Institute", Moscow 117218, Russia}
\affiliation{Argonne National Laboratory, Argonne, Illinois 60439}
\affiliation{American University of Cairo, New Cairo 11835, New Cairo, Egypt}
\affiliation{Brookhaven National Laboratory, Upton, New York 11973}
\affiliation{University of California, Berkeley, California 94720}
\affiliation{University of California, Davis, California 95616}
\affiliation{University of California, Los Angeles, California 90095}
\affiliation{University of California, Riverside, California 92521}
\affiliation{Central China Normal University, Wuhan, Hubei 430079 }
\affiliation{University of Illinois at Chicago, Chicago, Illinois 60607}
\affiliation{Creighton University, Omaha, Nebraska 68178}
\affiliation{Czech Technical University in Prague, FNSPE, Prague 115 19, Czech Republic}
\affiliation{Technische Universit\"at Darmstadt, Darmstadt 64289, Germany}
\affiliation{ELTE E\"otv\"os Lor\'and University, Budapest, Hungary H-1117}
\affiliation{Frankfurt Institute for Advanced Studies FIAS, Frankfurt 60438, Germany}
\affiliation{Fudan University, Shanghai, 200433 }
\affiliation{University of Heidelberg, Heidelberg 69120, Germany }
\affiliation{University of Houston, Houston, Texas 77204}
\affiliation{Huzhou University, Huzhou, Zhejiang  313000}
\affiliation{Indian Institute of Science Education and Research (IISER), Berhampur 760010 , India}
\affiliation{Indian Institute of Science Education and Research (IISER) Tirupati, Tirupati 517507, India}
\affiliation{Indian Institute Technology, Patna, Bihar 801106, India}
\affiliation{Indiana University, Bloomington, Indiana 47408}
\affiliation{Institute of Modern Physics, Chinese Academy of Sciences, Lanzhou, Gansu 730000 }
\affiliation{University of Jammu, Jammu 180001, India}
\affiliation{Joint Institute for Nuclear Research, Dubna 141 980, Russia}
\affiliation{Kent State University, Kent, Ohio 44242}
\affiliation{University of Kentucky, Lexington, Kentucky 40506-0055}
\affiliation{Lawrence Berkeley National Laboratory, Berkeley, California 94720}
\affiliation{Lehigh University, Bethlehem, Pennsylvania 18015}
\affiliation{Max-Planck-Institut f\"ur Physik, Munich 80805, Germany}
\affiliation{Michigan State University, East Lansing, Michigan 48824}
\affiliation{National Research Nuclear University MEPhI, Moscow 115409, Russia}
\affiliation{National Institute of Science Education and Research, HBNI, Jatni 752050, India}
\affiliation{National Cheng Kung University, Tainan 70101 }
\affiliation{Nuclear Physics Institute of the CAS, Rez 250 68, Czech Republic}
\affiliation{Ohio State University, Columbus, Ohio 43210}
\affiliation{Old Dominion University, Norfolk, VA 23529, USA}
\affiliation{Institute of Nuclear Physics PAN, Cracow 31-342, Poland}
\affiliation{Panjab University, Chandigarh 160014, India}
\affiliation{Pennsylvania State University, University Park, Pennsylvania 16802}
\affiliation{NRC "Kurchatov Institute", Institute of High Energy Physics, Protvino 142281, Russia}
\affiliation{Purdue University, West Lafayette, Indiana 47907}
\affiliation{Rice University, Houston, Texas 77251}
\affiliation{Rutgers University, Piscataway, New Jersey 08854}
\affiliation{Universidade de S\~ao Paulo, S\~ao Paulo, Brazil 05314-970}
\affiliation{University of Science and Technology of China, Hefei, Anhui 230026}
\affiliation{Shandong University, Qingdao, Shandong 266237}
\affiliation{Shanghai Institute of Applied Physics, Chinese Academy of Sciences, Shanghai 201800}
\affiliation{Southern Connecticut State University, New Haven, Connecticut 06515}
\affiliation{State University of New York, Stony Brook, New York 11794}
\affiliation{Instituto de Alta Investigaci\'on, Universidad de Tarapac\'a, Arica 1000000, Chile}
\affiliation{Temple University, Philadelphia, Pennsylvania 19122}
\affiliation{Texas A\&M University, College Station, Texas 77843}
\affiliation{University of Texas, Austin, Texas 78712}
\affiliation{Tsinghua University, Beijing 100084}
\affiliation{University of Tsukuba, Tsukuba, Ibaraki 305-8571, Japan}
\affiliation{United States Naval Academy, Annapolis, Maryland 21402}
\affiliation{Valparaiso University, Valparaiso, Indiana 46383}
\affiliation{Variable Energy Cyclotron Centre, Kolkata 700064, India}
\affiliation{Warsaw University of Technology, Warsaw 00-661, Poland}
\affiliation{Wayne State University, Detroit, Michigan 48201}
\affiliation{Yale University, New Haven, Connecticut 06520}

\author{J.~Adam}\affiliation{Brookhaven National Laboratory, Upton, New York 11973}
\author{L.~Adamczyk}\affiliation{AGH University of Science and Technology, FPACS, Cracow 30-059, Poland}
\author{J.~R.~Adams}\affiliation{Ohio State University, Columbus, Ohio 43210}
\author{J.~K.~Adkins}\affiliation{University of Kentucky, Lexington, Kentucky 40506-0055}
\author{G.~Agakishiev}\affiliation{Joint Institute for Nuclear Research, Dubna 141 980, Russia}
\author{M.~M.~Aggarwal}\affiliation{Panjab University, Chandigarh 160014, India}
\author{Z.~Ahammed}\affiliation{Variable Energy Cyclotron Centre, Kolkata 700064, India}
\author{I.~Alekseev}\affiliation{Alikhanov Institute for Theoretical and Experimental Physics NRC "Kurchatov Institute", Moscow 117218, Russia}\affiliation{National Research Nuclear University MEPhI, Moscow 115409, Russia}
\author{D.~M.~Anderson}\affiliation{Texas A\&M University, College Station, Texas 77843}
\author{A.~Aparin}\affiliation{Joint Institute for Nuclear Research, Dubna 141 980, Russia}
% ...

\collaboration{STAR Collaboration}\noaffiliation

\date{\today}

\begin{abstract}
We report results on the total and elastic cross sections in proton-proton collisions at $\sqrt{s}=200$~GeV obtained with the Roman Pot setup of the STAR experiment
at the Relativistic Heavy Ion Collider (RHIC).
The elastic differential cross section was measured in the squared four-momentum transfer range $0.045 \leq -t \leq 0.135$~GeV$^2$.
The value of the exponential slope parameter $B$ of the elastic differential cross section $d\sigma/dt \sim e^{-Bt}$ in the measured $-t$ range
was found to be $B = 14.32 \pm 0.09 (stat.)^{\scriptstyle +0.13}_{\scriptstyle -0.28} (syst.)$~GeV$^{-2}$. The total cross section $\sigma_{tot}$, obtained from extrapolation of the $d\sigma/dt$ to the optical point at $-t = 0$, is $\sigma_{tot} = 54.67 \pm 0.21 (stat.) ^{\scriptstyle +1.28}_{\scriptstyle -1.38} (syst.)$ mb.
We also present the values of the elastic cross section $\sigma_{el} = 10.85 \pm 0.03 (stat.) ^{\scriptstyle +0.49}_{\scriptstyle -0.41}(syst.)$ mb,
the elastic cross section integrated within the STAR $t$-range  $\sigma^{det}_{el} = 4.05 \pm 0.01 (stat.) ^{\scriptstyle+0.18}_{\scriptstyle -0.17}(syst.)$ mb, and the
inelastic cross section $\sigma_{inel} = 43.82 \pm 0.21  (stat.) ^{\scriptstyle +1.37}_{\scriptstyle -1.44} (syst.)$ mb.
The results are compared with the world data.
\end{abstract}


\maketitle


\section{Introduction}\label{sec:Intro}
Elastic scattering plays an important role in proton-proton ($pp$) scattering at high energies, as evidenced by the fact that it contributes about
20\% of the total cross section at the highest Large Hadron Collider (LHC) energies~\cite{bib:totem0}.
The $pp$ elastic and total cross sections have been measured at colliders with center of mass energies $ 2.76 \le \sqrt{s} \le 13$~TeV at the LHC~\cite{bib:totem0}
and at the Intersecting Storage Rings (ISR) at $\sqrt{s}=62.4$~GeV~\cite{Amaldi:1976zi}.
It is important, however, to have measurements in the energy gap between the ISR and the LHC to constrain the phenomenological models of the $pp$ cross sections since one still expects a difference between $pp$ and proton-antiproton ($p \bar p$) cross sections within the RHIC energy range.
The latter were measured up to $\sqrt{s}=1.8$ TeV at the Tevatron~\cite{bib:E710-88, bib:E710-89-1, bib:E710-89-2, bib:CDF, bib:E811, bib:D0}.
Both the values of the cross sections and the difference between $pp$ and $p \bar p$ affect phenomenological models~\cite{Bourrely:2002wr,Kopeliovich:2012yy,bib:COMPETE,bib:Barone,bib:DonnachieText,bib:Block2016}.

\section{The Experiment}\label{sec:ExpSetup}

The results presented here were obtained by the STAR experiment~\cite{bib:STARNIM} upgraded with the Roman Pot (RP) system used previously by the PP2PP experiment~\cite{Bultmann:2004ke}.
The current RP system was installed downstream of the STAR main detector at RHIC and was used to detect forward-scattered protons.
A modification of the vacuum chamber was required and the RP system was fully integrated with the STAR experiment.
With the addition of the RP system, the STAR physics program now includes $pp$ elastic scattering and two other measurements that require the detection of forward protons: Central Exclusive Production~\cite{Sikora:2018cyk}
 and particle production in both Single Diffraction Dissociation and Central Diffraction~\cite{bib:FulekDiff2018}. In these inelastic events, the components of the main part of the STAR detector are used to characterize the recoil system at central rapidity.

\begin{figure}[htb]
\includegraphics[width=0.45\textwidth]{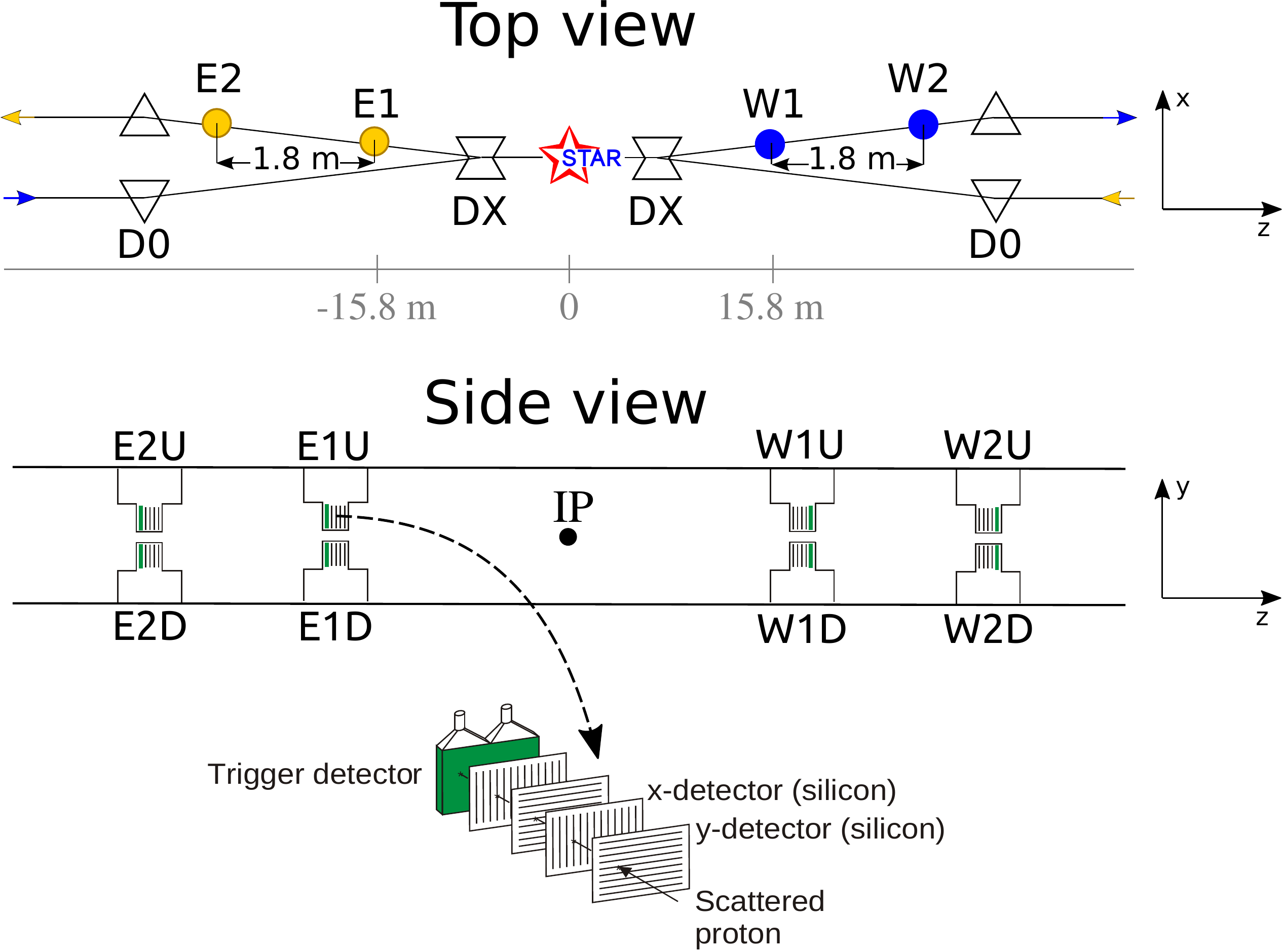}
\begin{center}
\caption{The layout of the experiment. The four Roman Pot stations (E1, E2) on the East side of STAR and (W1, W2) on the West side of STAR are shown. In the upper panel, the view in the $x,z$ plane is shown. In the lower panel, the $y,z$ view is shown with the detector package, which includes four Si strip detector planes and the trigger scintillation counter. Two dipole magnets DX and D0, which bend the beams into and out of the IP, are also shown.}
\label{fig:Figure1}
\end{center}
\end{figure}

The location of the RPs, top and side view, and the four Si detectors and a trigger scintillation counter package in each of the RPs are shown schematically in Fig.~\ref{fig:Figure1}.
The four planes of Si strip detectors~\cite{Bultmann:2004ke} with a pitch of 100 $\mu$m, two measuring the $x$-coordinate (X planes) and two measuring the $y$-coordinate (Y planes), were used to reconstruct the position of the proton at the RP.
The scintillation counter in each RP was used for triggering on candidate events with forward protons. It was read by two photomultiplier tubes (PMTs) for redundancy and high trigger efficiency. The trigger required at least one valid signal in at least one out of eight possible PMTs on each side of the interaction point (IP).

The location between DX and D0 RHIC dipole magnets is such that no special accelerator conditions such as large $\beta^*$ (the value of the betatron function at the IP) and parallel-to-point focusing, were needed to operate the RPs together with the rest of the STAR experiment's physics program.

The DX magnet and the detectors in the two RPs allow the measurement of the momentum vector of the scattered protons at the detection point.
Using the known bending angle of the DX magnet, one can determine the scattering angle in the $x,z$ plane, $\theta_x$.
Because of the symmetry of the RHIC rings, the field in the DX magnets on both sides of the IP are identical at the $10^{-3}$ level.
Hence, the bending angles of the magnets are also the same.
The scattering angle in the $y,z$ plane,  $\theta_y$, is determined from the $y$-coordinate  measured in the RPs. Consequently, the local angles at the RPs $\theta_x$, $\theta_y$ are the same as the scattering angles at the IP.

The data were acquired with normal $\beta^*=0.85$ m and were taken during the last four hours of an eight-hour store during the $pp$ run in 2015.
The last four hours were chosen to have beams with reduced tails, thus with lower singles rates and background in the RP trigger counters.
Three special luminosity measurements using Van der Meer scans~\cite{bib:VanDerMeer} were performed to determine the luminosity and to reduce
the systematic uncertainty on the luminosity measurement.
The RPs were moved as close to the beam as possible, to about $8\sigma_y$ of the beam size in the $y$-coordinate, which was closer than during nominal
data taking. The average instantaneous luminosity was  $\approx 45\cdot 10^{30}\:\text{cm}^{-2}\text{s}^{-1}$.
For this luminosity, the number of interactions per bunch crossing was 0.225 on average. Hence, pileup is not a concern.

There were about 6.7 million triggered events collected for the integrated luminosity of $1.8\: \text{pb}^{-1}$.
The closest position of the first readout strip was about 30 mm or about $10\rm{\sigma_{y}}$ of the beam,
which corresponds to a minimum $|t|$ of about 0.03 $\text{GeV}^2 $. The aperture of the DX magnet sets a maximum achievable limit of $|t| \approx 0.16 \text{ GeV}^2$, 
corresponding to a scattering angle of $\theta \approx 4 \text{ mrad}$.

\section{Alignment and Track Reconstruction}\label{sec:AlignmTrack}

Track reconstruction in the Si detectors was a three-step process: clustering that is used to determine the position of the proton trajectory in the Si plane, alignment to obtain the position of the proton in the elastic scattering coordinate system (the coordinate system in which two protons are collinear); and the reconstruction of a track, which leads to the reconstruction of the scattering angle needed to determine the $t$-value.

 \subsection{Clustering}\label{sec:Clustering}

To reconstruct track points in the RPs, we start with a clustering procedure for each Si detector plane separately.
In the first step, the noise cut that selects energies greater than $3\sigma_\text{{RMS}}$ above the pedestal is applied for each strip.
Then the clustering procedure searches for the channel with the maximum signal and a continuous series of channels adjacent to it.
This cluster is then removed from the pool of hits in a given plane, and the procedure is repeated until there are no more hits in the plane.
The position of the cluster is calculated as an energy-weighted average of the strip positions and their energies.
The energy distribution of reconstructed clusters is well described by the convolution of Landau and Gauss distributions.

To reconstruct the $x$-coordinate the positions of clusters found in both X planes were compared.
Given the limit on the maximum scattering angle of 4 mrad (Sec.~\ref{sec:ExpSetup}) and the distance $\Delta z = 14$ mm between two X planes,
a pair of clusters was accepted to calculate the $x$-coordinate if their position difference
$\Delta x$ satisfied condition that $ \Delta x \leq 2\cdot d_{strip} \approx 200$~$\mu$m, where $\text{d}_{strip}$ is the strip pitch.
The $x$-coordinate of the track was calculated as an average of the matched cluster positions.
The same procedure was done for $y$-coordinate using Y planes.
Positions of pairs of matched clusters found in the detector planes measuring the same coordinate define $x,y$ coordinates of space points for a given RP. In about 95\% of events, only one reconstructed space point in an RP was found.

\subsection{Alignment}\label{sec:Alignm}

Before the reconstruction of the scattering angle, an alignment procedure was performed in two steps, each producing one set of offsets. In the first step, survey data were utilized. That survey was done by the survey group of the accelerator department after the installation of the detector packages in the RPs. This survey determined the $x,y$ position of the first strip in each detector package with respect to the accelerator coordinate system. In the second step, corrections to the survey alignment were obtained using reconstructed elastic events with the constraint of collinearity of elastic scattering for tracks reconstructed on each side of
the IP.
To make sure that the sample consisted of the cleanest elastic events, it was also required that these two point tracks were uniquely reconstructed (one and only one reconstructed point in each RP), providing two track points on each side of the IP.

For each event, a least squares line fit was done to the four reconstructed points. Then, the mean
value of residuals  for each detector plane, which was the average distance of reconstructed points from the fitted line, was calculated.
Those mean residuals were used to correct the first strip position in each silicon detector plane, and the alignment process was then repeated with those new strip positions
until residual distributions were centered at zero, giving the optimal relative positions between RPs on opposite sides of each detector arm separately.
Typically three iterations were needed to obtain the offsets. The result of the second alignment step was a set of offsets in the coordinate system of the elastic scattering,
where two outgoing protons are collinear. Those offsets were used to correct the positions of the Si strips from which the scattering angles $\theta_x, \theta_y$
were reconstructed.

This alignment procedure was performed for each data run used in the analysis, and the mean value of run-by-run corrections was applied for each detector plane.
By its construction, the alignment offsets were obtained in the system of coordinates where two protons are elastically scattered, hence collinear (elastic scattering geometry).
Hence, the procedure left one variable unknown: the trajectory of the unscattered beam in the above coordinate system resulting from a beam-tilt angle in the collider,
which affects the $t$-scale of the differential distribution $dN/dt$. The procedure to estimate the beam-tilt angle is described in section~\ref{sec:Simul},
where Monte Carlo (MC) corrections are described.

\subsection{Scattering Angle and $t$ Reconstruction}\label{sec:AngleFourMom}

For small scattering angles $\theta$,  which are of the order of a few mrad, the positions of the track point $x_{RP},y_{RP}$  at a given RP station are given by:
\begin{equation}
 x_{RP} = x_{IP} + \theta_{x}(z_{RP} - z_{IP}) \mathrm{\hspace{0.2cm}}
 y_{RP} = y_{IP} + \theta_{y}( z_{RP} - z_{IP})
 \label{eq:PointPos}
\end{equation}
where $x_{IP}, y_{IP}, z_{IP} $ is the position of the primary vertex, $z_{RP}$ is the surveyed $z$-position of the RP station,
and $\theta_x$, $\theta_y$  are the scattering angles. Since the position of the primary vertex is not known on an event-by-event basis,
two reconstructed points are required to calculate the scattering angle. A track was defined by the two points reconstructed in the two detector stations on the same side of the IP.
The scattering angles $\theta_x$ and $\theta_y$ were determined by fitting a straight line using events with four track points, two on each side of the IP.
Given the beam momentum $p$ and small  scattering angles $\theta_x$ and $\theta_y$, the $t$-value was calculated using:
\begin{equation}
-t = (p_{in} - p_{out} )^2= p^2\theta^2 = p^2\cdot ( \theta_x^2 + \theta_y^2).
\label{eq:t-def}
\end{equation}

The  resolution in $t$, $\Delta t$, is dominated by the beam angular divergence, as given by the machine emittance and by the beta value at the collision point ($\beta^*$),
and to a much lesser extent by the detector resolution.
Thus, $\Delta t / t$ can be approximated by the term due to the beam angular divergence. For $p$ = 100~GeV and $\delta \theta$ = 175~$\mu$rad and taking into account averaging over the two beams the $\Delta t / t$ is given by:

\begin{equation}
\frac{\Delta t}{t} =  \sqrt{2} p \delta \theta / \sqrt{|t|} = 2.47\times 10^{-2} \text {GeV}/\sqrt{|t|}\label{eq:tresol}.
\end{equation}

\section{Data Analysis}\label{sec:DataAnal}

\begin{figure}[htbp]
\begin{center}
\includegraphics[width=0.5\textwidth]{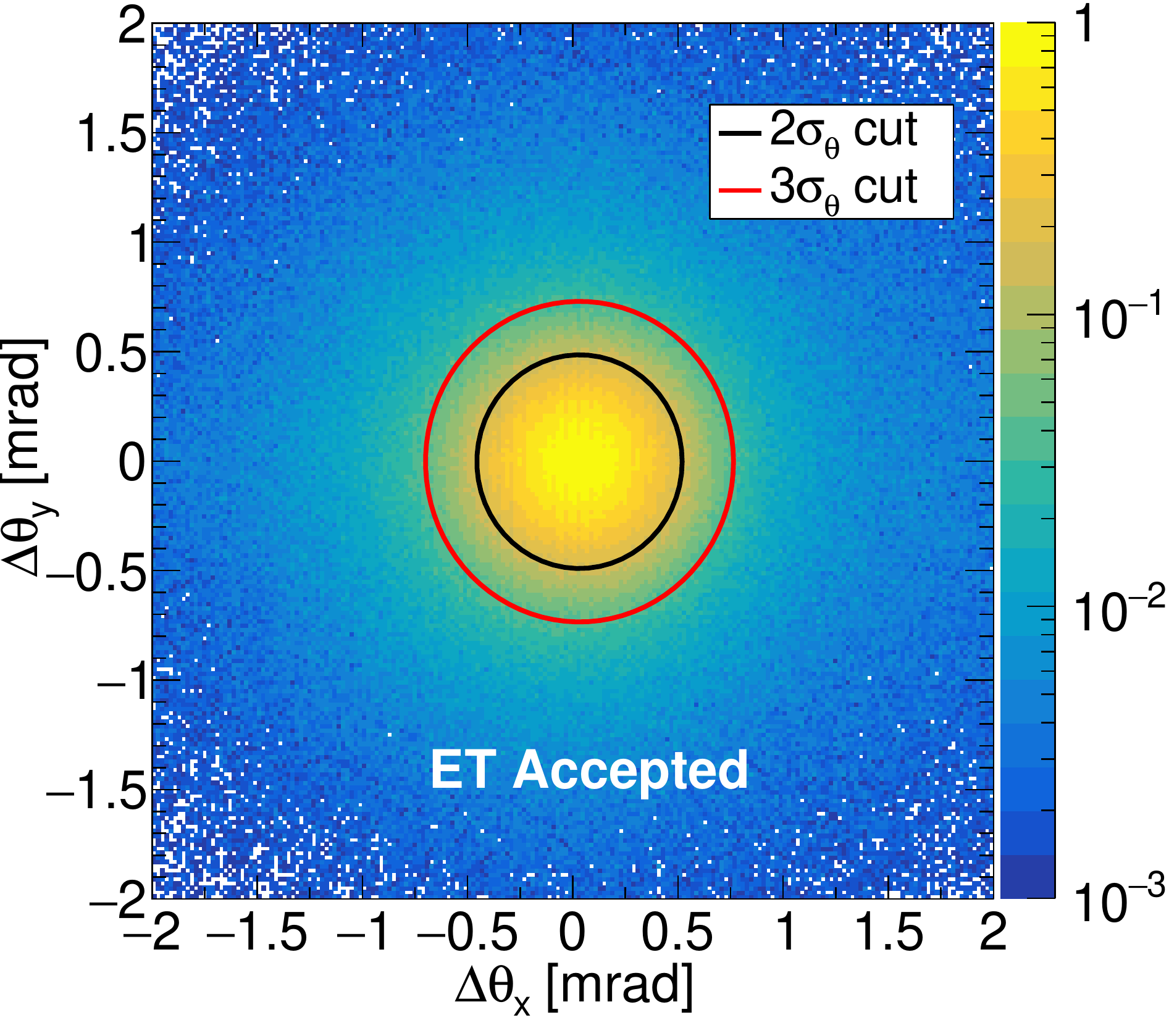}
\caption{Collinearity of the data sample $\Delta\theta_y$ vs $\Delta\theta_x$ for ET accepted events is shown. It is defined as the differences $\Delta\theta_x$ and $\Delta\theta_y$ between scattering angles $\theta_x,\theta_y$ reconstructed on the East and West side of the IP. It is plotted with the contours of $2\sigma_{\theta}$ and $3\sigma_{\theta}$,
 where $\sigma_{\theta} = 244 ~\mu\text{rad}$.}
\label{fig:Figure2}
\end{center}
\end{figure}

\begin{figure}[htbp]
\begin{center}
\includegraphics[width=0.45\textwidth]{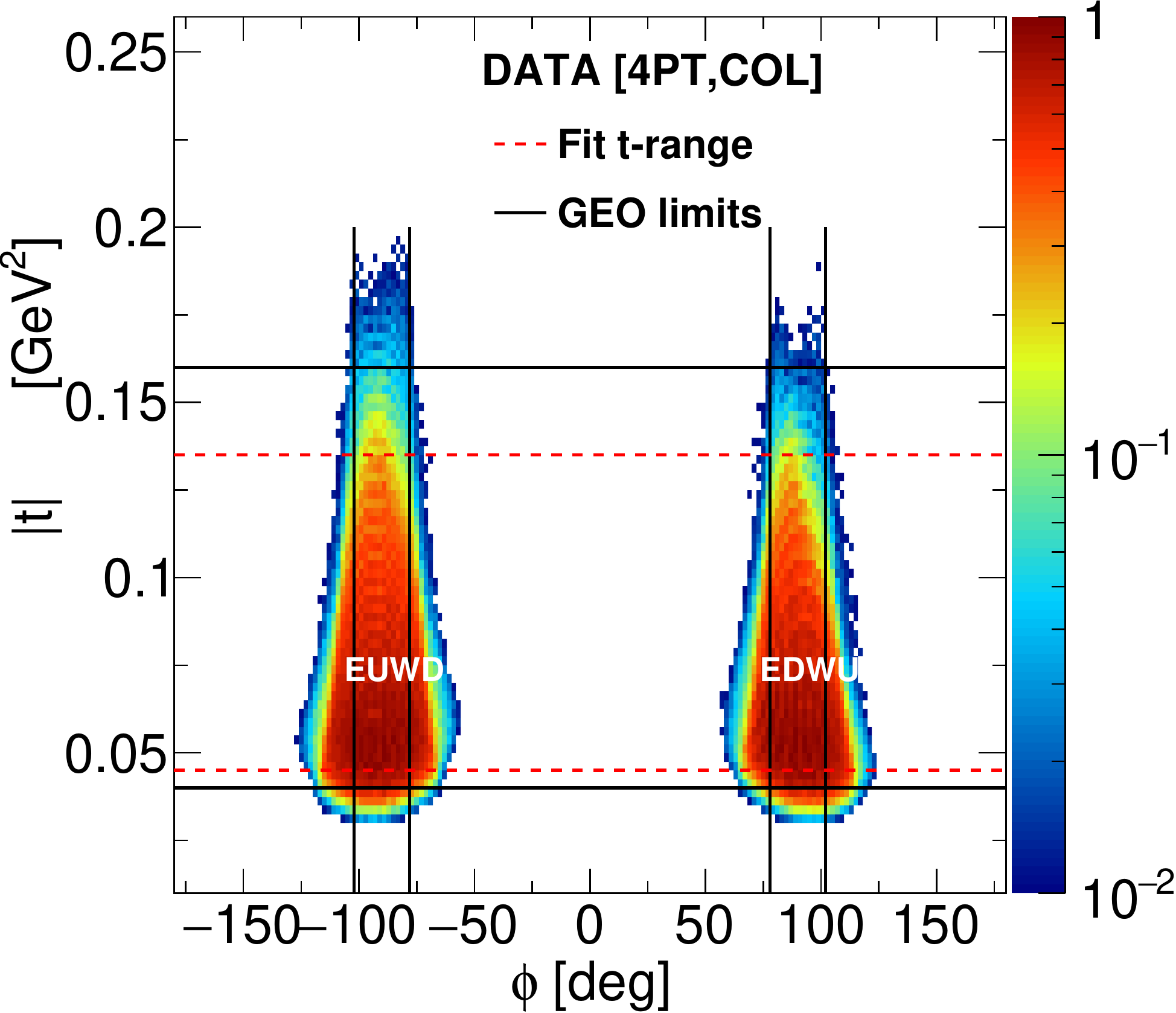}
\caption{Four-momentum transfer $|t|$ vs $\phi$ distributions for data for four-point collinear (4PT COL) events.
 The two elastic combinations of tracks between East and West, EUWD and EDWU, are shown. Each distribution is normalized to 1.}
\label{fig:Figure3}
\end{center}
\end{figure}

Because of the inclusive trigger condition, the collected data sample included the contributions from background, which consisted mostly of non-elastic events,
elastic protons scattered on the apertures and accidental coincidences of the beam halo. The basic feature of the elastic scattering is that the two outgoing protons are back to back. 
This is called a collinearity condition, which is used as a main selection criterion of elastic events. The following cuts were used to select clean elastic events from the collected data sample:

\begin{enumerate}
\item {\bf Elastic event topology (ET)}: Only events with a combination of reconstructed points in the RPs consistent with elastic scattering were accepted. Namely, the combinations with the lower East detector in coincidence with the upper West detector (EDWU), or the upper East detector in coincidence with the lower West detector (EUWD) have by definition the elastic event-hit pattern due to momentum conservation. In Fig.~\ref{fig:Figure2}, we show the collinearity condition $\Delta\theta_y$ vs $\Delta\theta_x$, 
where $\Delta\theta_x=\theta_x^W - \theta_x^E$ and $\Delta\theta_y=\theta_y^W - \theta_y^E$. 
Here, the $\theta_x^W, \theta_x^E, \theta_y^W, \theta_y^E$ are scattering angles reconstructed on the East and West sides of the IP, using the coordinates measured at the RP and the average IP position. 
The  contours of $2\sigma_{\theta}$ and $3\sigma_{\theta}$ are also shown. A clear peak of elastic events is seen.
\item {\bf 4-point track (4PT) data sample}: Only events with two-point tracks on the East and two-point tracks on the West (one track point in each RP in elastic combination) were kept.
\item {\bf Collinear (COL) events}: Since elastic events must satisfy a collinearity condition, collinearity in  $\theta^W$, $\theta^E$ was required.
Here, the $\theta^W$, $\theta^E$ are reconstructed scattering angles on the West and East sides of the IP. Since $\Delta\theta = \theta_{W} - \theta_{E} = 0$, collinearity within $2\sigma_{\theta}$ was required, namely $\Delta\theta < 2\sigma_{\theta}$, where $\sigma_{\theta} = 244 $ $\mu\text{rad}$ is the Gaussian width of the collinearity distribution, 
consistent with the beam angular divergence. 
The collinearity condition required also the radial distance between the two projected tracks in $x$ and $y$ at $z = 0$ to be within $5\sigma$ radius of the Gaussian width of its radial distance.
The $2\sigma_{\theta}$ cut was chosen to minimize background as described in Sec.~\ref{sec:Simul}.
\item {\bf Fiducial volume GEO cut}: After the elastic event candidates were chosen based on collinearity, one more set of cuts in a fiducial volume $(\phi, |t|)$, where $\phi$ is the azimuthal angle of the scattered proton, was needed to remove the remaining background. To stay away from the beam halo, the minimum $|t|$ corresponding to $12\sigma$ of the beam size was required; this was well outside of the beam envelope. Hence, the coincidence of the beam halo from the two beams is not expected.

To stay away from the apertures, additional cuts on maximum $|t|$ and $\phi$-range in ($\phi, |t|$) space were also required. 
They are shown in Fig. \ref{fig:Figure3}, where the lines labeled "GEO limits" show the limits of the geometrical acceptance and the fit range in ($\phi, |t|$) space accordingly. 
These cuts were chosen based on the simulation, which is described in Sec.~\ref{sec:Simul}. They were 
$78 < |\phi| < 102$ deg and $0.045 \leq -t \leq 0.135$~GeV$^2$.  

\end{enumerate}

We started with 6.607M events. After the ET cut there were 3.974M events left, 1.648M after the 4PT cut and 1.306M after the collinearity COL cut. The final sample had 0.666M events after the fiducial GEO cut.

\section{Simulation and correction factors}\label{sec:Simul}

Response of the detector was studied using a Monte Carlo data sample (G4MC) obtained with a GEANT4-based~\cite{bib:GEANT4} software package.
The simulation had a detailed implementation of the beam line and RP detector position, and of the Si detector readout behavior, where the point-reconstruction efficiency
in each RP was determined from the data.
The physics generator used for the simulation produced only elastic $pp$ scattering at $\sqrt{s} = 200$~GeV, as described by Eq.~\ref{eq:ethadronic},
namely $dN/dt \propto \exp{( -B|t|)}$ with $B = 14$~GeV$^{-2}$ and
uniform distribution in $\phi$. The kinematic range was $ -\pi \leq \phi \leq \pi $ and $0.01 \leq -t \leq 0.5$~GeV$^2$.
The simulation was used to correct the measured $dN/dt$ distributions from which the cross sections were obtained.

Using this simulation, the efficiency corrections were obtained as a function of $t$:
\begin{equation}
\epsilon (t_{reco}) = \frac{(dN/dt)^{MC}_{gen}}{(dN/dt)^{MC}_{reco}}
 \label{eq:CorrFactor}
\end{equation}
where $(dN/dt)^{MC}_{gen}$ and $(dN/dt)^{MC}_{reco}$ are the true and reconstructed distributions, respectively, based on a MC event sample
which passed reconstruction and selection steps identical to those applied to the experimental data. 
The $t_{reco}$ is the $t$-value calculated at the end of the MC reconstruction chain, using the same procedure as in the data analysis. 
The geometrical acceptance of the detector was the main contribution to the efficiency corrections.

The differential distribution $ (dN/dt)^{DATA} $ obtained from data was corrected using a ``bin-by-bin'' method according to Eq.~\ref{eq:CorrectionForm}
with  correction factors from Eq.~\ref{eq:CorrFactor}:
\begin{equation}
 \bigg ( \frac{dN}{dt} \bigg )^{DATA}_{corr} = \bigg ( \frac{dN}{dt} \bigg )^{DATA}_{reco} \times \epsilon(t_{reco}).
 \label{eq:CorrectionForm}
\end{equation}

Based on the MC simulation, the $(|t|,\phi)$ region of the acceptance for the $d\sigma/dt$ fit was chosen so that it had a slowly varying dependence 
on $|t|$, which is shown in in Fig.~\ref{fig:Figure5}.
\begin{figure}
\begin{center}
 \includegraphics[width=0.45\textwidth]{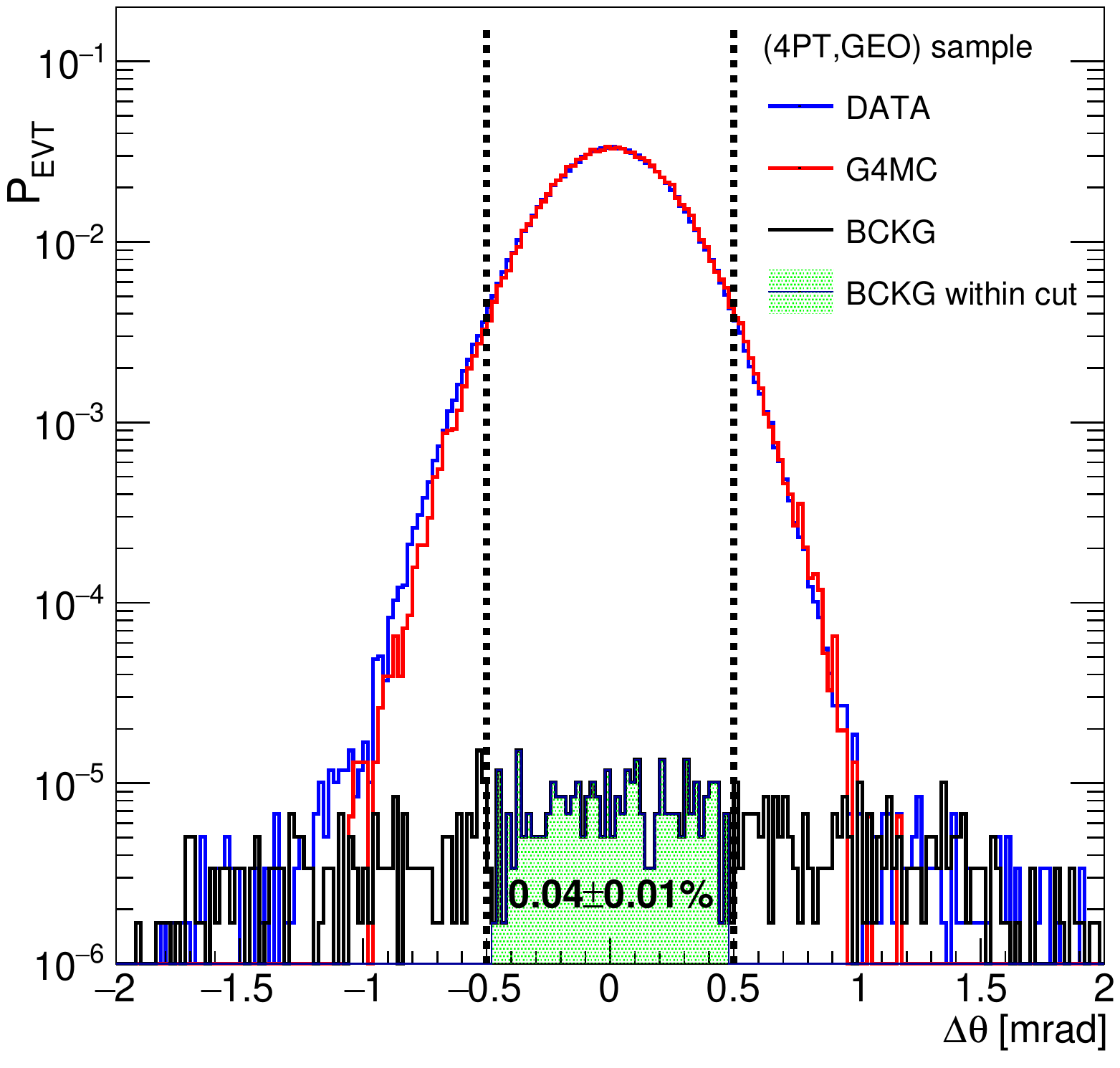}
 \caption{\small 
Collinearity, $\Delta\theta=\theta^W-\theta^E$, for data is compared with prediction from the G4MC MC. Both samples were required to pass fiducial volume cuts (4PT, GEO). Estimated background (BCKG), and background remaining after the collinearity cut (green area), are also shown. The vertical axis is probability per event (P$_{EVT}$).}
 \label{fig:Figure4}
 \end{center}
 \end{figure}

Additional corrections that needed to be considered were due to a possible non-zero initial colliding-beam angle (beam-tilt angle) and to the $x,y$ position of the beam at the IP in the coordinate system of reconstructed elastic events.
Such a beam tilt affects the $t$-scale of the measurement.
Note that the offset due to the $x,y$ position of the beam at the IP, being a parallel shift, does not change the reconstructed scattering angles $\theta_x, \theta_y$,  which are the result of fitting a straight line to the four-point events.

The beam-tilt angle causes offsets $\tau_{x}$ and $\tau_{y}$ of the reconstructed $\theta_{x}$ and
$\theta_{y}$ angles. This leads to an offset in the calculated $t$-values, which in lowest order is given by:
\begin{equation}
 \Delta t \simeq 2 \cdotp p^{2} \cdotp ( \theta_{x} \cdotp \tau_{x} + \theta_{y} \cdotp \tau_{y}).
\label{eq:DeltatTilt}
\end{equation}

Since the efficiency correction function was obtained from an MC simulation with a beam trajectory parallel to the detector local coordinate $z$-axis,
this beam-tilt angle needed to be accounted for in the MC simulated efficiency correction function.

To determine $\tau_x$ and $\tau_y$ the $dN/dt$ distribution from the data, Fig.~\ref{fig:Figure5} was used. The $\tau_x$, $\tau_y$ angles were varied within $[-0.2, 0.3]$ mrad and fitted to the data looking for the best fit probability. That best $\chi^2$ determined the beam crossing angles $\tau_x$ and $\tau_y$ to be 0.15 mrad and 0.015 mrad, respectively. 
Note that the $\tau_y$ is negligible compared to typical scattering angles of a few mrad. The contribution to the systematic uncertainties from the tilt angle was evaluated as described in Sec.~\ref{sec:Results}.

A GEANT4-based simulation was also used to study  protons interacting with material in front of the RPs such as the beam pipe, magnet structure and RF shield inside the DX-D0 chamber, etc.
In Fig.~\ref{fig:Figure4}, we compare the collinearity distributions for reconstructed data and reconstructed MC samples. We see a very good agreement between MC and the data. The vertical axis in Fig.~\ref{fig:Figure4} is the probability per event (P$_{EVT}$).  An estimate of the background (bckg) contribution is also shown. 
It was obtained using unpaired protons in the whole elastic trigger data sample by flipping the sign of $(x, y)$ coordinates of reconstructed points on one side of the IP. 
Then, the cuts of the analysis procedure were applied to all the events. 
This study is sensitive to the beam halo and to the inelastic events in our data sample.
Consequently, it made it possible to estimate the total (see Fig.~\ref{fig:Figure4}) and differential dN/dt background contribution. 
The latter was subtracted from the final $d\sigma/dt$, to estimate impact of the background on the fit results. 
We found small changes of $B$-slope and $d\sigma_{el} /dt \vert _{t=0}$, 0.006~GeV$^{-2}$ and -0.006~mb/GeV$^{2}$ respectively. 
These values were added in quadrature to the total systematic uncertainty. But given the number of significant digits, they did not change the result in Table~\ref{tab:FinSysErr}

Also, since the beam momentum uncertainty was at the $10^{-3}$ level, it was neglected. The RP point reconstruction efficiency implemented in the MC simulation was obtained from the data. The trigger efficiency determined from the data was essentially 100\%, so no corrections were made.

\section{Results}\label{sec:Results}

Over the  $t$-range of this measurement $0.045 \leq -t \leq 0.135$~GeV$^2$, the differential cross section $d\sigma/dt$ is dominated by the hadronic term, whose $t$-dependance is well described by an exponential with one free slope parameter $B$ and the normalization factor: 

\begin{equation}
\frac{d\sigma_{el}^{had}}{dt} = \left.\frac{d\sigma_{el}^{had}}{dt} \right|_{t=0}\cdot e^{-B|t|}
\label{eq:ethadronic}
\end{equation}

Hence, a two-parameter exponential fit was performed to the measured differential cross-section $d\sigma/dt$ to obtain the slope parameter $B$. We performed fitting using the bin center. 

The total cross section was obtained using the optical theorem, given in Eq.~\ref{eq:opth}, which relates the total cross section to the value of the hadronic elastic cross section at $t=0$:

\begin{equation}
{\sigma^2_{tot}} = { \left( \frac{ 16 \pi \left( {\hbar c} \right) ^{2} }{ { 1 + {\rho}^{2} } } \right) }
\left.\frac{d\sigma_{el}^{had}}{dt} \right|_{t=0}.
\label{eq:opth}
\end{equation}

The $\rho$ parameter in Eq.~\ref{eq:opth} is the ratio of the real to the imaginary part of the hadronic scattering amplitude and it was not measured in this experiment. 
Its value was obtained from a fit to the world data using the COMPETE~\cite{bib:COMPETE} model, which is based on Regge theory~\cite{bib:Barone,bib:DonnachieText}.
Because $\rho = 0.12$ and enters Eq.~\ref{eq:opth} in quadrature, the uncertainty on $\rho$ does not contribute significantly to the obtained value of $\sigma_{tot}$.
For the $\rho$-uncertainty we varied its value by $\pm 0.05$ and fitted Eq.~\ref{eq:ethadronic} to get the estimate of the corresponding systematic uncertainty.

\begin{figure}[htb]
\begin{center}
\includegraphics[width=0.45\textwidth]{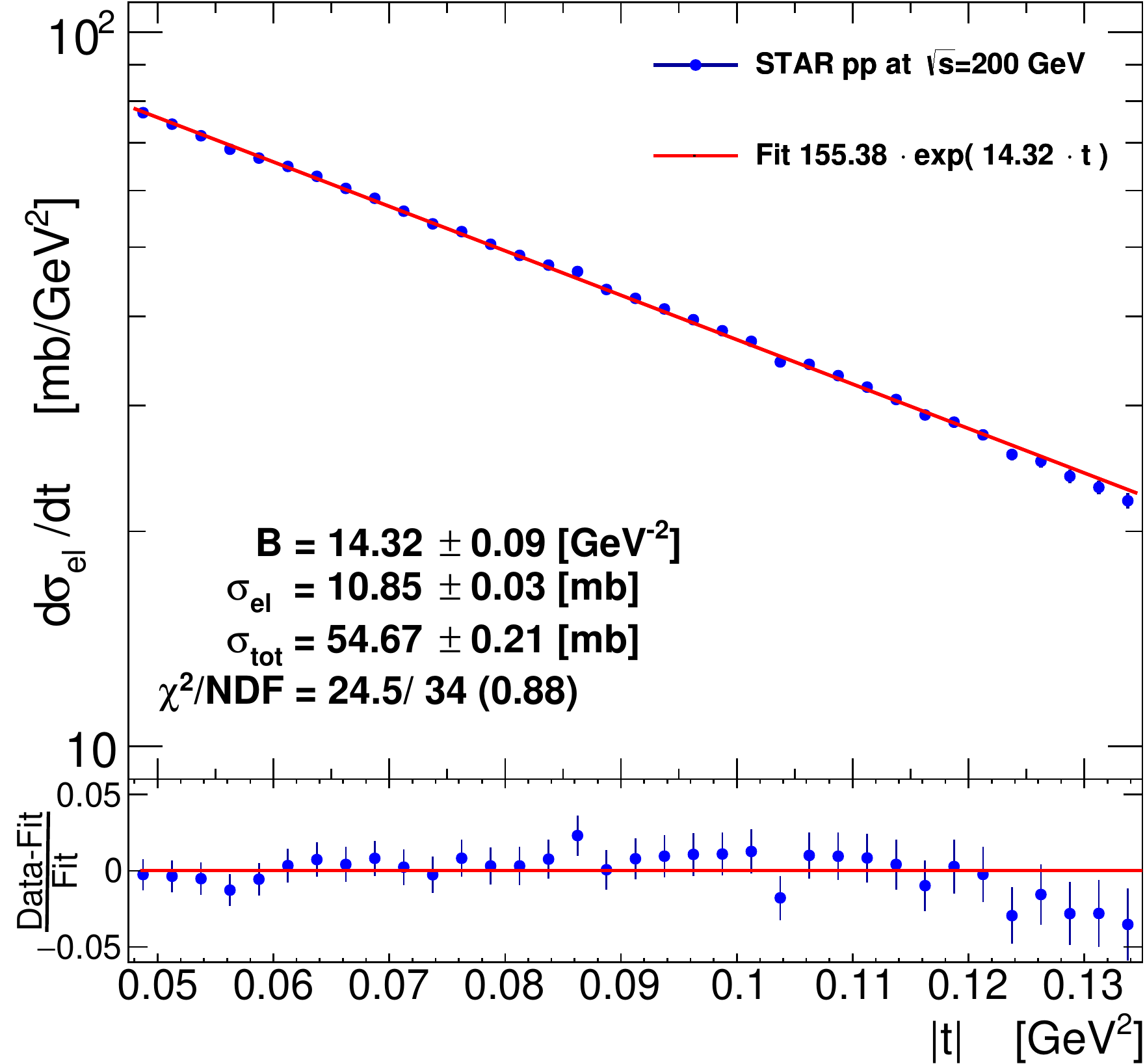}
 \caption{Top panel: $pp$ elastic differential cross-section $d\sigma/dt$ fitted with exponential $A\exp{(Bt)}$. 
The green triangles and the scale on the right-hand side of the plot show the MC simulated geometrical acceptance.
Bottom panel: Residuals (Data - Fit)/Fit. Uncertainties are statistical only.}
 \label{fig:Figure5}
 \end{center}
 \end{figure}

The fit of the Eq.~\ref{eq:ethadronic} with its results is shown in Fig.~\ref{fig:Figure5}.
The bin size in the fitted histogram is $0.0025$~GeV$^2$, which is smaller than the t-resolution. 
However, the fit was repeated with larger bin sizes by factor 2, 3 and 4 and also the MC based study of bin-to-bin migration showed that actual bin size does have a significant impact on  the fit parameter values except to increase statistical uncertainties with decreasing $NDF$ of the fit.

The dependence of the MC correction factors on the value of the initial slope $B$ was also investigated. The initial MC $t$-distributions were reweighed with the slope from the reconstructed data at detector level $B_{det}=14.8$~GeV$^{-2} $, and the correction factors were recalculated.  The fit results to $B$ and to $d\sigma_{el} /dt \vert _{t=0}$  changed by 0.01~GeV$^{-2}$ and 0.01~mb/GeV$^{2}$ respectively.
Since they did not change the total systematic uncertainty within accuracy displayed in Table~\ref{tab:FinSysErr}, they are not listed in there. 

The evaluation of the uncertainties due to the beam angular divergence, the vertex positions and their spread, and incoming beam angles was based
on MC simulations described in the previous section. We found that the largest single source of the systematic error of the $t$-scale of the experiment was due to the beam-tilt angle.
This shift of the $t$-distribution scale was studied with the MC simulation using the upper limits on the beam-tilt angle obtained from data. It resulted in an uncertainty on the fitted slope parameter of about 2\%.

We observe a weak dependence of the fitted slope $B$ and $\sigma_{tot}$ on the values of the beam-tilt angles, which were accounted for in a contribution to the systematic uncertainties.

For the cross section measurements, the largest systematic uncertainty is due to luminosity determination, which was estimated to be 4\%.
This is the scale uncertainty on the vertical scale of the cross section plot. It introduces a corresponding systematic uncertainty to the cross sections listed in Table~\ref{tab:FinSysErr}.

As  described in Sec.~\ref{sec:Simul}, the estimated background contribution due to the particle interactions with the material in front of the RPs and
within the geometrical acceptance used for this analysis was negligible, hence such a correction was not required.
\begin{table*}[httb]
 \centering
 \caption{ Results summary with systematic uncertainties. }
 \begin{tabular}{c|c|r|c|c|c|c|c}
 \hline \hline
 \multicolumn{3}{ c|}{Quantity}          & Statistical & \multicolumn{4}{|c}{ Systematic uncertainties}\\
 \cline{1-3} \cline{5-8}
 & & & & & & & \\ [-2ex]
  Name        & Units            & Value & Uncertainty & Beam tilt & Luminosity & $\rho$-parameter & Total sys. \\ [0.1ex]
 \hline
  & & & & & & \\ [-2.0ex]
 \dsigel0    & [mb/GeV$^{2}$]    & 155.38 & \mpm 1.19   & $^{+1.19}_{-0.91}$ & $^{+7.05}_{-6.47}$  &   $-$          & $^{+7.15}_{-6.53}$ \\ [1ex]
 \textbf{$B$}  & [GeV$^{-2}$]      &  14.32 & \mpm 0.09   & $^{+0.13}_{-0.28}$ &         $-$      &   $-$            & $^{+0.13}_{-0.28}$  \\ [1ex]
  $\sigma_{tot}$    & [mb]       &  54.67 & \mpm 0.21   & $^{+0.21}_{-0.64}$ & $^{+1.23}_{-1.15}$   & $^{+0.27}_{-0.41}$  & $^{+1.28}_{-1.38}$  \\ [1ex]
  $\sigma_{el}$     & [mb]       &  10.85 & \mpm 0.03   & $^{+0.07}_{-0.04}$ & $^{+0.49}_{-0.41}$   &   $-$            & $^{+0.49}_{-0.41}$  \\ [1ex]
  $\sigma^{det}_{el}$ & [mb]      &   4.05 & \mpm 0.01   & $^{+0.02}_{-0.01}$ & $^{+0.18}_{-0.17}$   &   $-$            & $^{+0.18}_{-0.17}$  \\ [1ex]
  $\sigma_{inel}$   & [mb]       &  43.82 & \mpm 0.21   & $^{+0.22}_{-0.64}$ & $^{+1.32}_{-1.22}$   & $^{+0.27}_{-0.41}$   & $^{+1.37}_{-1.44}$  \\ [0.5ex]
 \hline
 \end{tabular}
  \label{tab:FinSysErr}
 \end{table*}

Table~\ref{tab:FinSysErr} contains our final results and uncertainty estimates with the six observables listed in the left column. They are:
the intercept of the differential cross section \dsigel0 ; the slope parameter $B$; the total cross section  $\sigma_{tot}$ obtained using optical theorem;
the elastic cross section $\sigma_{el}$, which was  obtained by simply integrating the fitted exponential over all $t$; the elastic cross section integrated within the STAR $t$-range  $\sigma^{det}_{el}$; and the inelastic cross section $\sigma_{inel}$, which was obtained by subtracting $\sigma_{el}$ from $\sigma_{tot}$.
As such, both  $\sigma_{el}$  and  $\sigma_{inel}$ are estimates.
Nevertheless, we see good agreement with the world data. This is because most of the  $\sigma_{el}$ is in the purely exponential region measured in this experiment.
The last column of Table~\ref{tab:FinSysErr} lists the total systematic uncertainty, which was obtained by adding the individual systematic uncertainties in quadrature.
The $\rho$-parameter column in the table lists the systematic uncertainty due to the uncertainty in the ratio of the real to the imaginary part of the hadronic scattering
amplitude. 

\begin{figure}
\centering
\includegraphics[width=0.45\textwidth]{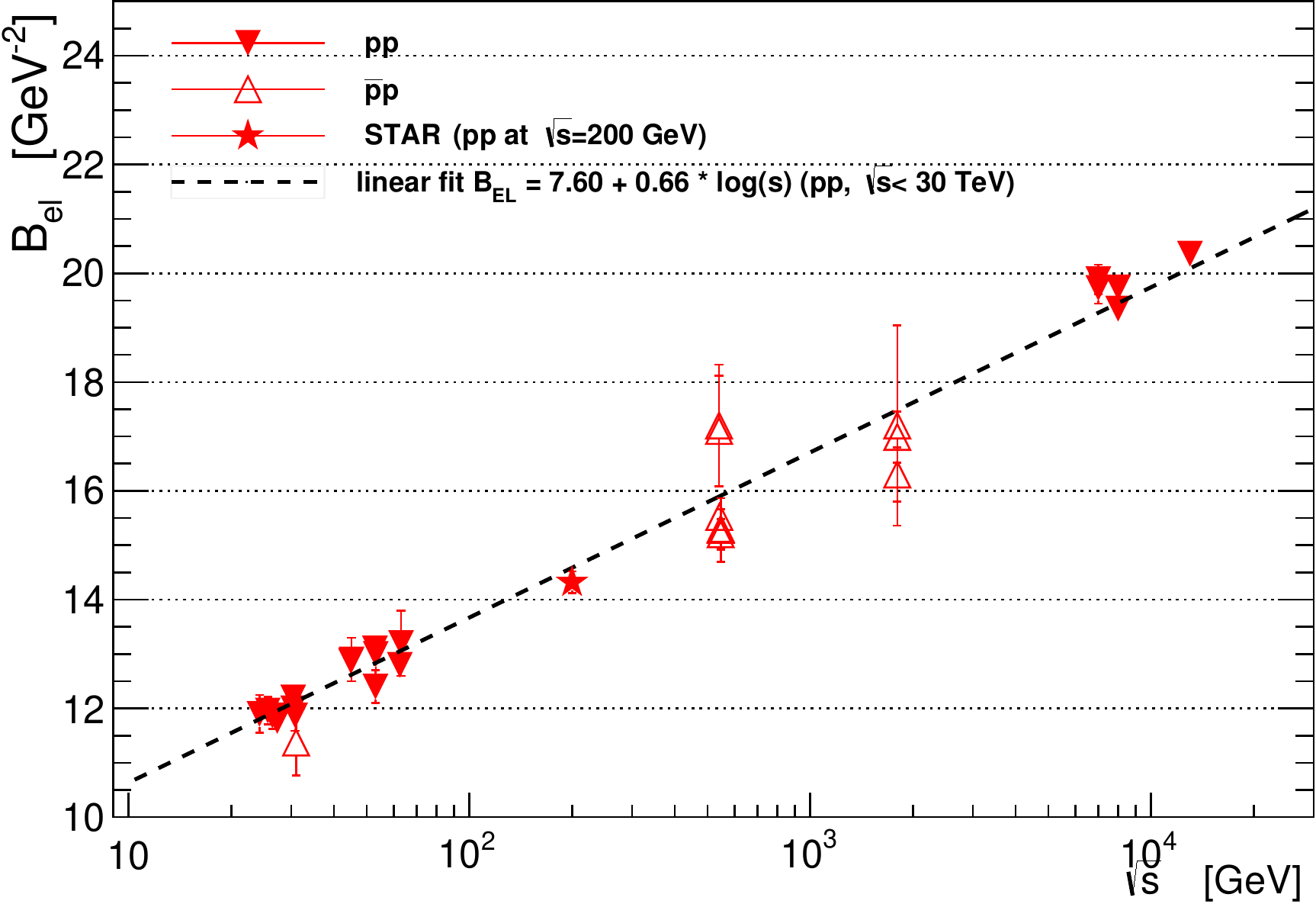}
\caption{Comparison of STAR result on $B$-slope with the world data with the $t$-range of this experiment. Below 1.8 TeV data are from~\cite{bib:HEPDurham},  the Tevatron data are \cite{bib:E710-88, bib:E710-89-1, bib:E710-89-2} and the LHC data are \cite{bib:totem0, bib:totem3, bib:totem4, bib:totem6, bib:atlas1, bib:atlas2}. The $t$-range for the world data was chosen to be compatible with the STAR $t$-range.}
\label{fig:Figure6}
\end{figure}
The asymmetric systematic uncertainties on the cross sections are due to the luminosity uncertainty, which is the dominant uncertainty of the measurement.

\begin{figure}
\centering
\includegraphics[width=0.45\textwidth]{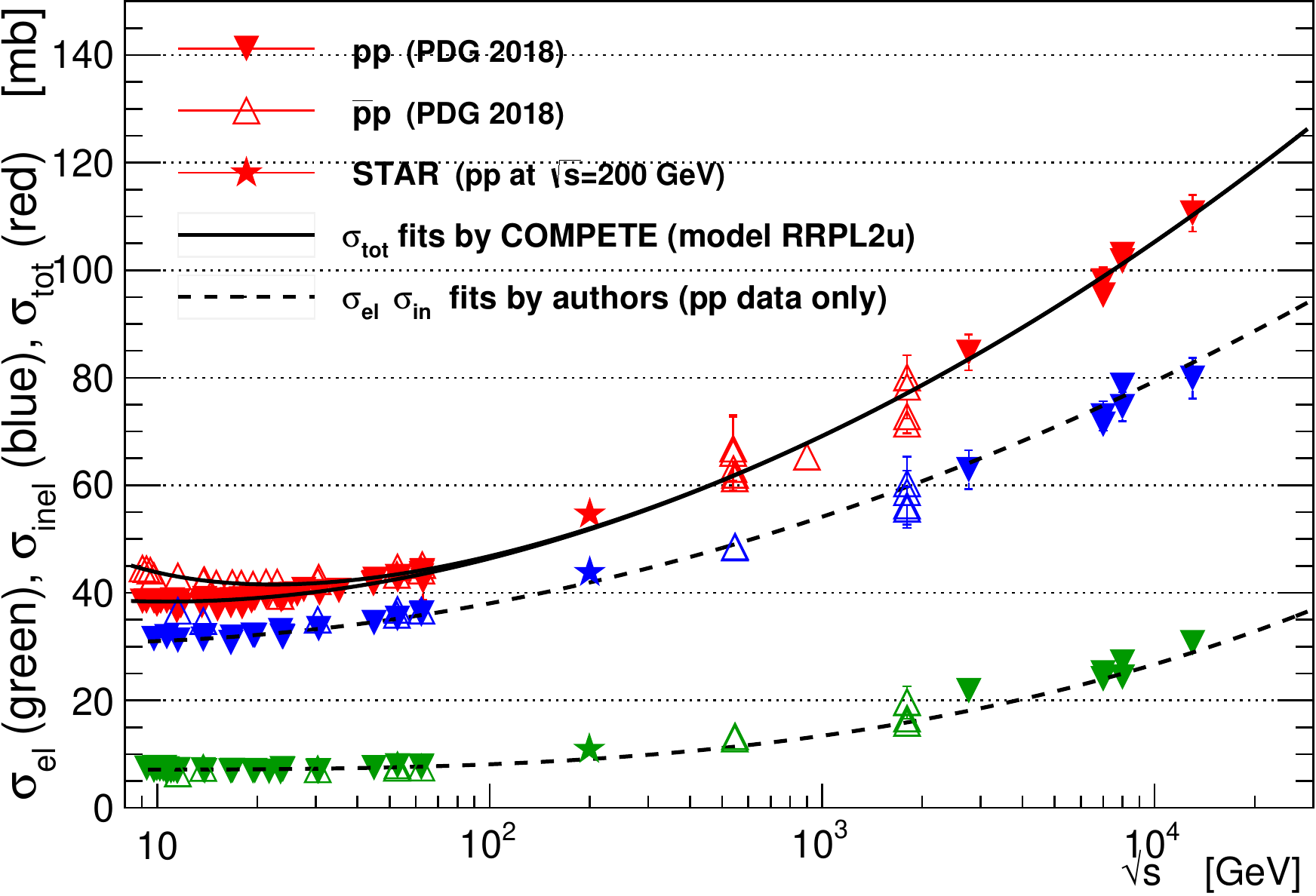}
\caption{Comparison of STAR results on $\sigma_{tot}$, $\sigma_{inel}$ and $\sigma_{el}$ with the world data for data below 1.8 TeV \cite{bib:PDG}, the Tevatron \cite{bib:E710-89-1, bib:E710-89-2, bib:CDF, bib:E811} and the LHC experiments \cite{bib:totem0, bib:totem3, bib:totem4, bib:totem5, bib:atlas1, bib:atlas2}. The COMPETE prediction for $\sigma_{tot}$ is also shown.
The dashed curves, represent STAR fits to $\sigma_{inel}$ and $\sigma_{el}$ using the same function as used by COMPETE. STAR data points were not used in the fit.}
\label{fig:Figure7}
\end{figure}

The comparison of our results with the world data on the nuclear slope parameter $B$ is shown in Fig.~\ref{fig:Figure6},
and on $\sigma_{tot}$, $\sigma_{inel}$, $\sigma_{el}$ are shown in Fig.~\ref{fig:Figure7}, where the total uncertainty of the STAR data points was obtained by adding the statistical and systematic uncertainties in quadrature. STAR results agree well with the world data and with the COMPETE  model~\cite{bib:COMPETE}, which is a fit to the existing world data available prior to this measurement and which is now commonly used as a reference comparison with the data.
\section{Summary}\label{sec:Summary}
The STAR experiment measured the elastic differential cross-section in $pp$ scattering as a function of $t$ in the range $ 0.045 \leq -t \leq 0.135$~GeV$^2$ at $\sqrt{s} = 200$~GeV.
This cross-section is well described by $e^{-B|t|}$ with the slope $B = 14.32 \pm 0.09(stat.)^{\scriptstyle +0.13}_{\scriptstyle -0.28}(syst.)$~GeV$^{-2} $.
The total $pp$ cross-section was found to be $\sigma_{tot} = 54.67 \pm 0.21(stat.)^{\scriptstyle +1.28}_{\scriptstyle -1.38}(syst.)$ mb.
Extrapolation of the measured differential elastic cross-section to the outside of the STAR $t$-acceptance permitted the determination
of $\sigma_{el} = 10.85 \pm 0.03(stat.)^{\scriptstyle +0.49}_{\scriptstyle -0.41}(syst.)$ mb. We also determined the elastic cross section
integrated within the STAR $t$-range  $\sigma^{det}_{el} = 4.05 \pm 0.01(stat.)^{\scriptstyle+0.18}_{\scriptstyle -0.17}(syst.)$ mb.
By subtracting the calculated $\sigma_{el}$ from $\sigma_{tot}$, we also obtained an inelastic cross section $\sigma_{inel} = 43.82 \pm 0.21(stat.)^{\scriptstyle +1.37}_{\scriptstyle - 1.44}(syst.) $ mb.
 We find that the obtained results are in good agreement with the world data. The $\sigma_{tot}$ agrees with the COMPETE prediction at $\sqrt s = 200$ GeV of 51.79 mb within about $2\sigma$ of the total uncertainty.

\vspace{0.1in}

{\bf Acknowledgements:}
We thank the RHIC Operations Group and RCF at BNL, the NERSC Center at LBNL, and the Open Science Grid consortium for providing resources and support.  This work was supported in part by the Office of Nuclear Physics within the U.S. DOE Office of Science, the U.S. National Science Foundation, the Ministry of Education and Science of the Russian Federation, National Natural Science Foundation of China, Chinese Academy of Science, the Ministry of Science and Technology of China and the Chinese Ministry of Education, the National Research Foundation of Korea, Czech Science Foundation and Ministry of Education, Youth and Sports of the Czech Republic, Hungarian National Research, Development and Innovation Office, New National Excellency Programme of the Hungarian Ministry of Human Capacities, Department of Atomic Energy and Department of Science and Technology of the Government of India, the National Science Centre of Poland, the Ministry  of Science, Education and Sports of the Republic of Croatia, RosAtom of Russia and German Bundesministerium fur Bildung, Wissenschaft, Forschung and Technologie (BMBF) and the Helmholtz Association.

\end{document}